\documentstyle[preprint,tighten,pra,aps,latexsym,amssymb]{revtex}

\begin{document}
\tightenlines
\draft

\preprint{UTF 422}

\title{Quantum Scalar Fields on Anti-de Sitter Spacetime}

\author{Marco M.~Caldarelli\footnote{email: caldarel@science.unitn.it}\\ 
\vspace*{0.5cm}}

\address{Universit\`a  degli Studi di Trento,\\ 
Dipartimento di Fisica,\\ 
Via Sommarive, 14\\ 
38050 Povo (TN)\\ 
Italia\\ 
\vspace*{0.5cm}      
and\\ Istituto Nazionale di Fisica Nucleare,\\ 
Gruppo Collegato di Trento,\\ Italia}

\maketitle
\begin{abstract}
We investigate the propagation of arbitrarily coupled scalar fields on the
$N$-dimensional hyperbolic space ${\mathbb H}^N$. Using the
$\zeta$-function regularization we compute exactly the one loop effective
action. The vacuum expectation value of quadratic field fluctuations and the
one loop renormalized stress tensor are then computed using the recently
proposed direct  $\zeta$-function  technique. Our computation tests the
validity of this approach in presence of a continuous spectrum. 
Our results apply as well to the $N$-dimensional anti-de~Sitter spacetime,
whose appropriate euclidean section is the hyperbolic space $ {\mathbb H}^N$.
\end{abstract}


\maketitle

\section{Introduction}

There has always been interest in anti-de~Sitter (AdS) spacetime and quantum
fields propagating on it. Being a maximally symmetric spacetime, it has
been an excellent model to investigate questions of principle related to
the quantization of fields propagating on curved background, the
interaction with the gravitational field and the issues related to its lack
of global hyperbolicity \cite{Fronsdal65,AIS78,BL85,IO85}.\\ 
The importance of this theoretical work increased when it was realized that
AdS spacetime emerges as a stable ground state solution of gauge extended
supergravity \cite{BF82} and Kaluza-Klein theories, in various dimensions.
Stability was also established for gravity fluctuations about the AdS
background \cite{AD82}.
Recently, there has been a revival of the interest in AdS spacetimes,
due to the AdS/Conformal field theory-correspondence conjecture
\cite{Maldacena97} and its relevance in the study of the large-$N$ limit of
nonabelian gauge theories.\\ 
Moreover, due to the negative cosmological constant, black holes with
nonspherical topology can be constructed on AdS background
\cite{ABHP96,Huang,Mann97,Vanzo97,LZ96,CZ96,KMV98,CJS98,Birmingham98}.
Some of them have a constant curvature and can be obtained as quotients of
AdS by a discrete subgroup of its isometry group, $SO(N-1,2)$; the most
popular are the Ba\~nados-Teitelboim-Zanelli solutions in three dimensions
\cite{BTZ92}, but higher dimensional generalizations exist
\cite{ABHP96,Banados97}. 
These black holes are locally isometric to AdS, and quantum corrections
due to the propagation of scalar fields on the background of these black
holes have been considered by various authors, for the BTZ black hole
\cite{LO94,Steif94,MZ97,BMVZ98}, for the singular background of toroidal
black holes in four dimensions \cite{Caldarelli98}, while the propagation
of photons in topological black hole spacetimes has been investigated in
\cite{Cai98}.\\ 
In this paper we shall study the propagation of a scalar quantum field,
with arbitrary coupling, on AdS spacetime, in the framework of euclidean
field theory. The appropriate euclidean section of AdS spacetime is the
hyperbolic space $ {\mathbb H}^N$ \cite{Camporesi91,DC76b}. We shall
compute the exact expressions at one loop of the effective action, the vacuum
expectation value of the field fluctuations and the renormalized stress
tensor in arbitrary dimension.  We shall use the powerful formalism of
$\zeta$-function renormalization \cite{Hawking77,DC76a,BCVZ96}. In particular,
the field fluctuations and the stress tensor will be computed with the
recently proposed direct  $\zeta$-function  approach \cite{MI98,Moretti97}. The
equivalence of this approach with the more standard (euclidean)
point-splitting procedure has been shown for compact spaces
\cite{Moretti98a,Moretti98b}, and holds only formally in the noncompact
case.
We shall deal with operators in $ {\mathbb H}^N$ with continuous spectrum,
and our results are an important test of the generalization of this equivalence
when the hypothesis of compactness misses.\\ 
This paper is organized as follows. In Section \ref{ha} we study some
spectral properties of Laplace-like operators in the hyperbolic spaces
$ {\mathbb H}^N$ and compute the related  $\zeta$-function . In Section
\ref{ea} we compute the one loop effective action.
The vacuum expectation value of the field fluctuations and of the
renormalized stress tensor are computed using the direct  $\zeta$-function
approach, in Sections \ref{sff} and \ref{st} respectively. In Section
\ref{app} we apply our formulae in various dimensions. We end drawing some
conclusions in Section \ref{concl} and with an Appendix, where 
an integral is analysed.

\section{Spectral analysis on $ {\mathbb H}^N$}\label{ha}

The $N$-dimensional anti-de~Sitter spacetime (AdS$_N$) with radius $a$ is
the hyperboloid \cite{HE73}
\begin{equation}
x_1^2+\cdots+x_{N-1}^2-u^2-v^2=-a^2
\end{equation}
embedded if flat $N+1$ dimensional spacetime $ {\mathbb R}^{N-1}_2$. It is
an homogeneous, constant curvature manifold. This spacetime solves
Einstein's equations with negative cosmological constant $\Lambda$ and
curvature scalar $R$ given respectively by
\begin{equation}
\Lambda=-\frac{(N-1)(N-2)}{2a^2},\qquad R=-\frac{N(N-1)}{a^2}.
\end{equation}
The definition of a quantum field theory on this manifold requires some care,
and has been extensively studied. The main problem comes from the fact that
it is not globally hyperbolic, and boundary conditions must be
supplemented. The choice of the boundary conditions is not unique, and
different unequivalent Fock representations exist \cite{AIS78}.\\ 
Furthermore, the correct euclidean formulation of a field theory is not
immediate, as the choice of the euclidean section on which to work is
ambiguous.
It is now an established fact that the appropriate euclidean section for
AdS$_N$ spacetime is the $N$-dimensional hyperbolic space ${\mathbb H}^N$. 
The analytic continuation to the euclidean section automatically selects a
particular representation for the quantum fields, corresponding to
Dirichlet boundary conditions at infinity \cite{Camporesi91,DC76b}. In the
following we shall restrict ourselves to the euclidean theory, where the
$\zeta$-function  renormalization technique is available.\\
We shall study the propagation of a scalar field $\phi(x)$ on 
${\mathbb H}^N$. Its action is given by
\begin{equation}
I[\phi]=- {1\over 2}\int \left(\nabla^\mu\phi\,\nabla_{\mu}\phi+m^2\phi^2
+\xi R\phi^2\right)\sqrt{g(x)}\,d^Nx,
\label{action}\end{equation}
where $m$ is the mass of $\phi$ and $\xi$ determines the coupling with the
scalar curvature $R$. The associated motion operator is
$-\Delta_{ {\mathbb H}^N}+m^2+\xi R$. It is an elliptic Laplace-like
operator, hence the first step is to study the spectral properties of the
laplacian.\\
It is convenient to work in the Poincar\'e half-space model of 
${\mathbb H}^N$; the coordinates are $(y,{\bf x})$, where $y>0$ is a ``radial''
coordinate and ${\bf x}\in {\mathbb R}^{N-1}$ parametrizes the flat
$(N-1)$-dimensional transverse manifold. In these coordinates the metric
is
\begin{equation}
ds^2=\frac{a^2}{y^2}\left(dy^2+dx_1^2+\cdots+dx_{N-1}^2\right)
\end{equation}
and the Laplace operator on $ {\mathbb H}^N$ reads
\begin{equation}
\Delta_{ {\mathbb H}^N}=\frac1{a^2}\left[y^2
\left(\partial_y^2+\Delta_{N-1}\right)-(N-2)y\partial_y\right],
\end{equation}
where $\Delta_{N-1}$ is the Laplacian on the flat $(N-1)$-dimensional
transverse manifold.
Let us define $\rho_N=(N-1)/2$ and the operator
$L=-\Delta_{ {\mathbb H}^N}-\rho_N^2$.
The eigenvalue equation for $L$, $a^2L\psi=\lambda^2\psi$, can be solved by
separation of variables, with the ansatz
\begin{equation}
\psi_{\lambda,{\bf k}}(x)=\phi_\lambda(y)f_{\bf k}({\bf x}).
\end{equation}
On the transverse manifold the eigenvalue equation becomes
\begin{equation}
-\Delta_{N-1}f_{\bf k}({\bf x})={\bf k}^2f_{\bf k}({\bf x}),\qquad
f_{\bf k}({\bf x})=(2\pi)^{-\rho_N}e^{i{\bf k}\cdot{\bf x}},
\end{equation}
and one gets for $\phi_\lambda(y)$,
\begin{equation}
\phi_\lambda''-\frac{N-2}y\phi'_\lambda+
\left(\frac{\lambda^2+\rho_N^2}{y^2}-{\bf k}^2\right)\phi_\lambda=0.
\end{equation}
This is a Bessel equation, and requiring the solutions to be well-behaved
at infinity, one gets $\phi_\lambda(y)=y^{\rho_N}K_{i\lambda}(ky)$, where
$K_\nu(x)$ is a Mc~Donald function and $k=|{\bf k}|$. As a result, the
spectrum is continuous and the generalized eigenfunctions are
\begin{equation}
\psi_{\lambda,{\bf k}}(x)=y^{\rho_N}K_{i\lambda}(ky)f_{\bf k}({\bf x}).
\label{autof}\end{equation}
The associated spectral measure, defined by
$(\phi_\lambda,\phi_{\lambda'})={\delta(\lambda-\lambda')}/{\mu(\lambda)}$,
is
\begin{equation}
\mu(\lambda)=\frac{2\lambda}{\pi^2}\sinh\pi\lambda.
\end{equation}
Hence the spectral theorem yields
\begin{equation}
\left<x|F(L)|x'\right>=\int_0^\infty\!\!d\lambda\,\mu(\lambda)
F(\lambda^2/a^2)\int\!\!\frac{d^{N-1}k}{(2\pi)^{2\rho_N}}(yy')^{\rho_N}
e^{i{\bf k}\cdot{\bf u}}K_{i\lambda}(ky)K_{i\lambda}(ky'),
\label{sf}\end{equation}
where ${\bf u}={\bf x}-{\bf x'}$. The spectral measure $\mu(\lambda)$ is
related to the $N$-dimensional Plancherel measure $p_N(\lambda)$, that
arises from the spectral analysis on the hyperboloid model of
${\mathbb H}^N$, by
\begin{equation}
\Gamma(\rho_N+i\lambda)\Gamma(\rho_N-i\lambda)\mu(\lambda)=
\frac{2^{2N-3}}{\pi^2}p_N(\lambda).
\label{pmu}
\end{equation}
The Plancherel measure is given by
\begin{equation}
p_N(\lambda)=\beta_N\prod_{j=0}^{\rho_N-1}(\lambda^2+j^2),
\end{equation}
for $N\geq 3$ odd, and by
\begin{equation}
p_N(\lambda)=\beta_N\lambda\tanh(\pi\lambda)\prod_{j=\frac 12}^{\rho_N-1}
(\lambda^2+j^2),
\end{equation}
for $N$ even (for $N=2$ the product is omitted). The coefficients $\beta_N$
are defined by
\begin{equation}
\beta_N=\frac\pi{2^{2(N-2)}\left[\Gamma\left(\frac N2\right)\right]^2}.
\end{equation}
It is useful to decompose the Plancherel measure in a sum of monomials,
with coefficients $c^N_n$, according to
\begin{equation}
p_N(\lambda)=\beta_N\sum_{n=1}^{\rho_N}c_{2n}\lambda^{2n},\qquad
p_N(\lambda)=\beta_N\tanh(\pi\lambda)\sum_{n=0}^{\frac N2-1}c_{2n+1}
\lambda^{2n+1},
\end{equation}
for $N$ odd and $N$ even respectively.\\ 
The relevant motion operator for the scalar field theory (\ref{action}) is
$L_b=L+b$, where we have defined the coupling parameter
$a^2b=\rho_N^2+a^2m^2+\xi a^2R$.
The operator $L_b$ is positive definite as long as $a^2b>0$, that is 
\begin{equation}
\xi < \frac{N-1}{4N}+\frac{a^2m^2}{N(N-1)} \equiv \xi_{crit};
\end{equation}
for $\xi>\xi_{crit}$ the ground state becomes unstable and the theory is
not defined, while for $\xi=\xi_{crit}$ there is a continuous spectrum
starting from zero and our procedure cannot be applied (not straightfully
at least).\\
The local  $\zeta$-function  associated to the operator $L_b$ acting on $
{\mathbb H}^N$ can be obtained from the spectral theorem, using formula
(\ref{sf}) with $F(x)=(x+b)^{-s}$, yielding
\begin{equation}
\zeta(s,x|L_b)={2^{N-3}\Gamma\left(\frac N2\right)\over\pi^{\frac N2+1}}
a^{2s-N}\int_0^\infty{p_N(\lambda)\,d\lambda\over(\lambda^2+a^2b)^s};
\end{equation}
note that the integrated  $\zeta$-function  coincides with the local
$\zeta$-function, as ${\mathbb H}^N$ is a homogeneous manifold.
We are interested in the meromorphic structure of the $\zeta$-function,
dictated by the integral
\begin{equation}
I_N(s)=\int_0^\infty{p_N(\lambda)\,d\lambda\over(\lambda^2+a^2b)^s}.
\end{equation}
It is convenient here to split the Plancherel measure in a sum of monomials
in $\lambda$ with coefficients $c^N_n$, and distinguish two cases according
to the parity of $N$. For odd $N$, the integral can be computed explicitly
in term of Euler's gamma function, yielding
\begin{equation}
I_N(s)={\sqrt\pi\beta_N\over2\Gamma(s)}\sum_{n=1}^{\rho_N}{(2n-1)!!\over 2^n}
c^N_{2n}(a^2b)^{n+\frac12-s}\Gamma\left( s-n-\frac12\right),
\label{INodd}\end{equation}
and the odd-dimensional  $\zeta$-function  reads
\begin{equation}
\zeta(s,x|L_b)={a^{2s-N}\over2^N\pi^{\rho_N}\Gamma\left(\frac N2\right)
\Gamma(s)}\sum_{n=1}^{\rho_N}{(2n-1)!!\over 2^n}
c^N_{2n}(a^2b)^{n+\frac12-s}\Gamma\left( s-n-\frac12\right).
\label{Zodd}
\end{equation}
This function can be analytically continued to a meromorphic function with
simple poles in $s=\frac N2-k$, with $k\in {\mathbb N}$. If $N$ is odd, the
$\tanh(\pi\lambda)$ factor in the Plancherel measure complicates a little
bit the computation; to show the analytic structure of $I_N(s)$ we shall
split the integral in two according to the relation
\begin{equation}
\tanh(\pi\lambda)=1-\frac2{e^{2\pi\lambda}+1}.
\end{equation}
We obtain
\begin{equation}
I_N(s)=\beta_N\sum_{n=0}^{\frac N2-1}c^N_{2n+1}\left(
 {1\over 2} n!(a^2b)^{n+1-s}{\Gamma(s-n-1)\over\Gamma(s)}
 -2H_n(s;a\sqrt b)\right),
\label{INeven}\end{equation}
where we have defined the function
\begin{equation}
H_n(s;\mu)=\int_0^\infty\frac{\lambda^{2n+1}\,d\lambda}
{(e^{2\pi\lambda}+1)(\lambda^2+\mu^2)^s}.
\label{Bn}\end{equation}
The exponential at denominator makes $H_n(s;\mu)$ an analytic function on the
whole complex $s$-plane. Some properties of this function are examined in
the Appendix; in particular $H_n(0;\mu)$ can be exactly calculated in terms of
Bernouilli numbers.\\ 
From (\ref{INeven}) we obtain the  $\zeta$-function  in an even-dimensional
hyperbolic space
\begin{equation}
\zeta(s,x|L_b)={a^{2s-N}\over2^{N-1}\pi^{\frac N2}\Gamma\left(\frac N2\right)}
\sum_{n=0}^{\frac N2-1}c^N_{2n+1}\left(
 {1\over 2} n!(a^2b)^{n+1-s}{\Gamma(s-n-1)\over\Gamma(s)}
 -2H_n(s;a\sqrt b)\right).
\label{Zeven}\end{equation} 
Hence the meromorphic structure of the $\zeta$-function, shared with
$I_N(s)$, is again completely dictated by the Euler's gamma functions; it
can be analytically continued to a meromorphic function with simple poles
located in $s=1,2,\dots,N/2$.\\ 
An important observation is that, in both cases, the  $\zeta$-function  is
well-defined in $s=0$, and it is hence possible to proceed with the
$\zeta$-function regularization.

\section{Effective Action for scalar fields}\label{ea}

In a path integral approach, the effective action for a scalar field can be
formally expressed as the functional determinant of the operator
$L_b$ as
\begin{equation}
I_{eff}=- {1\over 2}\ln\det(L_b/\mu^2),
\label{Ieff}
\end{equation}
where $\mu$ is an arbitrary renormalization mass scale coming from the
path-integral measure. This determinant is however a formally divergent
quantity and needs to be regularized. We shall proceed here with the
$\zeta$-function renormalization. In this framework, the regularized
determinant reads
\begin{equation}
\ln\det(L_b/\mu^2)=-\zeta'(0|L_b)-\zeta(0|L_b)\ln\mu^2.
\end{equation}
Let us handle first the odd dimensional case. First of all, we note that
$\zeta(0|L_b)=0$ in odd dimensions, and the dependence from the
renormalization scale drops out. To compute the derivative of the
$\zeta$-function in $s=0$, we note that it consists in a sum of terms of the
form $f(s)/\Gamma(s)$, with $f(s)$ a smooth function of $s$, and that
\begin{equation}
\lim_{s\rightarrow 0}{d\over ds}\left(\frac{f(s)}{\Gamma(s)}\right)=f(0).
\end{equation}
Computing the derivative term by term in Eq. (\ref{Zodd}) we easily obtain
\begin{equation}
\ln \det\left( L_b/\mu^2\right)=\frac{a^{-N}}{2^N\pi^{\frac N2-1}\Gamma
\left(\frac N2\right)}
\sum_{n=1}^{\rho_N}\frac{(-1)^nc^N_{2n}}{n+\frac12}(a^2b)^{n+\frac12}
\end{equation}
for $N$ odd.\\ 
Let us turn now to the case of even dimensionality. Now the
$\zeta$-function  does not vanish anymore, and we have to keep the
renormalization scale. This time we have to deal with a sum of functions of
the form
\begin{equation}
F_n(s)={\Gamma(s-n-1)\over\Gamma(s)}=\prod_{k=1}^{n+1}\frac1{s-k}
\end{equation}
that assume in $s=0$ the value 
\begin{equation}
F_n(0)=\frac{(-1)^{n+1}}{(n+1)!};
\end{equation}
inserting it into Eq. (\ref{Zeven}) we obtain $\zeta(0|L_b)$. The computation
of $\zeta'(0|L_b)$ can be done using the relation
\begin{equation}
F'_n(0)=\frac{(-1)^{n+1}}{(n+1)!}\sum_{k=1}^{n+1}\frac1k,
\end{equation}
and, with a bit of algebra, one obtains the effective action on an
even-dimensional hyperbolic space
\begin{eqnarray}
\ln \det\left( L_b/\mu^2\right)=\frac{a^{-N}}{2^{N-1}\pi^{N/2}\Gamma
\left(\frac N2\right)}
\sum_{n=0}^{\frac N2-1}c^N_{2n+1}\left[
{(-1)^n(a^2b)^{n+1}\over2(n+1)}\left( d_{n+1}-\ln(b/\mu^2)\right)
\right.\nonumber\\ 
\left.+2H'_n(0;a\sqrt b)+2H_n(0)\ln\left( a^2\mu^2\right)
\right],
\end{eqnarray}
where we have defined for convenience
\begin{equation}
d_0=0,\qquad d_n=\sum_{k=1}^n\frac1k \quad (n\geq1).
\end{equation}
As expected, the arbitrary mass scale $\mu$ combines with the radius $a$
and the coupling parameter $b$ to leave a dimensionless argument for the
logarithm.

\section{Vacuum expectation value of the Field Fluctuations}\label{sff}

The vacuum expectation value of the field fluctuations can be computed
within the  $\zeta$-function  regularization scheme by means of the formula
\cite{MI98}
\begin{equation}
\left<\phi^2(x)\right>=\left.{d\over ds}\right|_{s=0}{s\over\mu^2}
\zeta(s+1,x|L_b/\mu^2)
=\lim_{s\rightarrow0}\left[(1+s\ln\mu^2)\zeta(s+1,x|L_b)+s\zeta'(s+1|L_b)
\right],
\label{ff}\end{equation}
where $\zeta(s,x|L_b)$ is the local  $\zeta$-function  and $\mu$ is again the
renormalization mass scale. Recently it has been shown that this procedure
leads to the same results as the point-splitting technique \cite{Moretti98a}.
The odd-dimensional case is simpler, as the local  $\zeta$-function  and
its derivative are finite in $s=1$: the field theory is
super-$\zeta$-regular and the field fluctuations are simply given by the
value in $s=1$ of the local $\zeta$-function.
From (\ref{Zodd}) we easily obtain the expectation value of the field
fluctuations of a scalar field in an odd-dimensional hyperbolic space
\begin{equation}
\left<\phi^2(x)\right> = {a^{2-N}\over{2^N\pi^{\frac N2-1}\Gamma
\left(\frac N2\right)}}
\sum_{n=1}^{\rho_N}(-1)^nc^N_{2n}(a^2b)^{n-\frac12}.
\label{ffodd}\end{equation}
The even-dimensional case has to be handled more carefully, because the
associated local  $\zeta$-function  has a pole in $s=1$. However, the poles
cancel exactly in Eq. (\ref{ff}) as we shall see.
The appearance of the poles is due to the presence of the function $F_n(s)$ 
in the local $\zeta$-function. Near $s=1$, $F_n(s)$ and its derivative
behave as
\begin{equation}
F_n(1+s)=\frac{(-1)^n}{n!}\frac1s+{\cal O}(s),\qquad
F'_n(1+s)=-\frac{(-1)^n}{n!}\left(\frac1{s^2}-\frac{d_n}s\right)+{\cal O}(s).
\end{equation}
Using these expressions, the behaviour of the local  $\zeta$-function  and
its derivative near $s=1$ is a simple matter of algebra, and one obtains
\begin{eqnarray}
\zeta(s+1,x|L_b)&=&\frac{a^{2-N}}{2^{N-1}\pi^{N/2}\Gamma\left(\frac N2\right)}
\sum_{n=0}^{\frac N2-1}c^N_{2n+1}\left[
 {1\over 2}(-a^2b)^n\frac1s-2H_n(1;a\sqrt b)\right]+{\cal O}(s),\\ 
\zeta'(s+1,x|L_b)&=&\frac{a^{2-N}}{2^{N-1}\pi^{N/2}\Gamma\left(\frac N2\right)}
\sum_{n=0}^{\frac N2-1}c^N_{2n+1}\left[-
  {1\over 2}(-a^2b)^n\left(\frac1{s^2}-\left( d_n-\ln b\right)
    \frac1s\right)\right.\nonumber\\ 
&&\left.-2H_n(1;a\sqrt b)\ln a^2-2H'_n(1;a\sqrt b)\right]+{\cal O}(s).
\end{eqnarray}
Inserting these expressions in (\ref{ff}), we see that the poles disappear and
the limit $s\rightarrow 0$ is smooth, yielding the expectation value for
the field fluctuations in an even-dimensional hyperbolic space 
\begin{equation}
  \left<\phi^2(x)\right>_{\mu^2}=\frac{a^{2-N}}{2^{N-1}\pi^{N/2}
  \Gamma\left(\frac N2\right)}\sum_{n=0}^{\frac N2-1}c^N_{2n+1}\left[
  {1\over 2}(-a^2b)^n\left( d_n-\ln(b/\mu^2)\right)-2H_n(1;a\sqrt b)\right].
\label{ffeven}\end{equation}
Again, the coupling parameter $b$ and the renormalization scale $\mu$
combine to leave a dimensionless argument for the logarithm. 

\section{One loop renormalized Stress Tensor}\label{st}

Finally, we turn to the computation of the renormalized stress tensor for a
quantum scalar field propagating in ${\mathbb H}^N$. It is possible to
perform the computation directly in the framework of the  $\zeta$-function
regularization \cite{Moretti97}.
In this approach, one defines the analytic continuation of the tensor
\begin{equation}
\zeta_ {\mu\nu}(s|L_b)(x)=\sum_n\lambda_n^{-s}T_ {\mu\nu}[\phi^*_n,\phi_n](x),
\label{st1}
\end{equation}
in which $\phi_n$ are the eigenfunctions of the Laplace-like operator $L_b$,
and $T_ {\mu\nu}[\phi^*_n,\phi_n](x)$ is the classical stress tensor evaluated
on the modes, defined as
\begin{equation}
T_ {\mu\nu}[\phi^*,\phi](x)=\frac2{\sqrt g}\frac{\delta I[\phi^*,\phi]}
{\delta g_ {\mu\nu}(x)},
\label{st2}
\end{equation}
$I[\phi^*,\phi]$ being the associated classical action.
Then, the vacuum expectation value of the stress tensor reads
\begin{eqnarray}
\left<T_ {\mu\nu}(x)\right>&=&\lim_{s\rightarrow 0}\left[
\zeta_{\mu\nu}(s+1,x|L_b)
+\frac12g_ {\mu\nu}\zeta(s,x|L_b)\right.\nonumber\\ 
&&\left.\phantom{\frac12}+s\left[\zeta'_{\mu\nu}(s+1,x|L_b)+
\zeta_{\mu\nu}(s+1,x|L_b)\ln\mu^2\right]\right].
\label{st3}
\end{eqnarray}
This limit is smooth; the computation is simplified observing that
\cite{Moretti97}
\begin{equation}
\zeta_ {\mu\nu}(s,x|L_b)=\bar\zeta_{\mu\nu}(s,x|L_b)
+L_{\mu\nu}\zeta(s,x|L_b)-{1\over 2}g_{\mu\nu}\zeta(s-1,x|L_b),
\label{st4}
\end{equation}
where we have defined the operator
\begin{equation}
L_ {\mu\nu}=-\xi\nabla_\mu\nabla_\nu+\left(\xi-\frac14\right)
g_{\mu\nu}\Delta+\xi R_ {\mu\nu},
\label{st5}
\end{equation}
and $\bar\zeta_ {\mu\nu}(s,x|L_b)$ is the analytical continuation of the series
\begin{equation}
\bar\zeta_ {\mu\nu}(s,x|L_b)= {1\over 2}\sum_n\lambda_n^{-s}
\left(\nabla_\mu\phi_n^*\nabla_\nu\phi_n+\nabla_\nu\phi^*_n
\nabla_\mu\phi_n\right).
\label{st6}
\end{equation}
The equivalence of the direct  $\zeta$-function  approach to the
computation of the one loop renormalized stress tensor with the
point-splitting approach has been shown in \cite{Moretti98b}.
Note that the proof of this equivalence has been carried out for compact
spaces, where the spectrum is discrete. Here we are dealing with a
continuous spectrum, and we shall check that this equivalence holds on
hyperbolic spaces.\\ 
The continuous spectrum generalization of Eq. (\ref{st6}) reads, 
making use of the eigenfunctions (\ref{autof}),
\begin{equation}
\bar\zeta_ {\mu\nu}(s,x|L_b)=\frac{a^{2s}}2\int_0^\infty
\!\!d\lambda\,\mu(\lambda)\int_{ {\mathbb R}^{N-1}}\!\!d^{N-1}k
\,(\lambda^2+a^2b)^{-s}
\left(\nabla_\mu\phi_n^*\nabla_\nu\phi_n
+\nabla_\nu\phi^*_n\nabla_\mu\phi_n\right).
\end{equation}
This integral can be carried out without big difficulties, yielding
\begin{eqnarray}
\bar\zeta_ {\mu\nu}(s,x|L_b)&=&\frac{\Gamma(\rho_N+1)a^{2s-2}}
{2\pi^{\rho_N}N!}g_ {\mu\nu}^{ {\mathbb H}^N}(x)\,
\int_0^\infty\!\!\!\Gamma(\rho_N+i\lambda)\Gamma(\rho_N-i\lambda)
\frac{\lambda^2+\rho_N^2}{(\lambda^2+a^2b)^s}\mu(\lambda)\,d\lambda\nonumber\\ 
&=&\frac{2^{N-3}a^{2s-2}\Gamma\left(\frac N2\right)}{\pi^{\frac N2+1}N}
g_{\mu\nu}^{{\mathbb H}^N}(x)\,\int_0^\infty\!\!\!
\frac{\lambda^2+\rho_N^2}{(\lambda^2+a^2b)^s}p(\lambda)\,d\lambda,
\end{eqnarray}
where we have used the relation (\ref{pmu}) between the Plancherel measure
and the measure $\mu(\lambda)$. We recognise in the last integral the
function $I_N(s)$. Now the tensor $\zeta_ {\mu\nu}(s,x|L_b)$ follows from Eq.
(\ref{st4}),
\begin{equation}
\zeta_ {\mu\nu}(s,x|L_b)=\frac{2^{N-3}\Gamma\left(\frac N2\right)}{\pi^{\frac
    N2+1}N}a^{2s-N}
\left[\left( 1-\frac N2\right) a^{-2}I_N(s-1)-m^2I_N(s)
\right]g_ {\mu\nu}^{ {\mathbb H}^N}(x).
\end{equation}
If the dimension $N$ of the hyperbolic space is odd, the function $I_N(s)$,
given in (\ref{INodd}), is finite in $s=0$ and $s=1$, where it assumes the
values $I_N(0)=0$ and 
\begin{equation}
I_N(1)= {1\over 2}\pi\beta_N\,\sum_{n=1}^{\rho_N}\,(-1)^nc^N_{2n}(a^2b)
^{n- {1\over 2}};
\end{equation}
the theory is hence super-$\zeta$-regular and there are no divergent terms
in (\ref{st3}), that leads to the vacuum expectation value of the stress
tensor in odd dimensions
\begin{equation}
  \label{Todd}
  \left<T_ {\mu\nu}(x)\right>^{ {\mathbb H}^N}_{\mu^2}=
  {m^2a^{2-N}\over 2^N\pi^{N/2-1}N\Gamma\left(\frac N2\right)}
  \left(\sum_{n=1}^{\rho_N}(-1)^{n+1}c^N_{2n}(a^2b)^{n-\frac12}
  \right)g_ {\mu\nu}^{{\mathbb H}^N} (x).
\end{equation}
In the even-dimensional case, the computation is more delicate as divergent
  terms appear in Eq. (\ref{st3}); however they cancel and the limit can be
  performed without excessive difficulty, leading to the following vacuum
  expectation value for the stress tensor in an even-dimensional hyperbolic
  space,
\begin{eqnarray}
  \label{Teven}
  \left<T_ {\mu\nu}(x)\right>_{\mu^2}^{ {\mathbb H}^N}&=&
  {a^{-N}\over 2^{N-1}\pi^{N/2}N\Gamma\left(\frac N2\right)}
  \,g_ {\mu\nu}^{{\mathbb H}^N} (x)
  \sum_{n=0}^{\frac N2-1}c^N_{2n+1}\left[
    \frac{(-1)^{n+1}}{2(n+1)}(a^2b)^{n+1}
  \right.\nonumber\\ 
  &&\left.
    +\frac{(-1)^{n+1}}2a^2m^2(a^2b)^n\left( d_n-\ln(b/\mu^2)\right)
    +2a^2m^2H_n(1;a\sqrt b)-2H_n(0)
  \right].
\end{eqnarray}
Correctly, the renormalization scale $\mu$ combines with the coupling
parameter $b$ to give a dimensionless argument for the logarithm.\\ 
We always find a stress tensor proportional to the metric, as
expected for an homogeneous manifold; this property implies that it is
automatically conserved.\\ 
We shall now proceed with some checks of this result. First of all there is
a general formula relating the vacuum expectation value of the field
fluctuations with the trace of the renormalized stress tensor \cite{MI98}
\begin{equation}
\left<T^\mu{}_\mu(x)\right>=\zeta(0,x|L_b)-\left[m^2+
\frac{\xi-\xi_N}{4\xi_N-1}\Delta\right]\left<\phi^2(x)\right>,
\label{relffst}\end{equation}
where $\xi_N=(N-2)/4(N-1)$ is the conformal coupling parameter
\footnote{The coefficient $1/2\xi_D$ which appears in (13) of \cite{MI98}
  is misprinted and has to be replaced by $1/(4\xi_D-1)$. See also
  Theorem~2.4. of \cite{Moretti98b}.}. 
Noting that
$\Delta\left<\phi^2(x)\right>=0$, it is easy to check that this relation is
verified, both in odd and even dimensions. $ {\mathbb H}^N$ being a homogeneous
manifold, the stress tensor is completely determined by its trace, and
$\left<T_ {\mu\nu}\right>$ can be computed directly from the field
fluctuations. We proceeded however to a direct  $\zeta$-function
computation to verify its validity in presence of a continuous spectrum.\\ 
Another check can be done in the conformally coupled case. This is defined
by $m=0$ and $\xi=\xi_N$, that is $a^2b=1/4$ for any dimension. In odd
dimension the stress tensor (\ref{Todd}) is proportional to $m^2$ and the
conformal anomaly correctly vanishes, while in even dimension we get,
taking the trace of (\ref{Teven}),
\begin{equation}
\left<T^\mu{}_\mu(x)\right>={a^{-N}\over 2^{N-1}\pi^{N/2}\Gamma
  \left(\frac N2\right)}
  \sum_{n=0}^{\frac N2-1}c^N_{2n+1}\left[
    \frac{(-1)^{n+1}}{2(n+1)}4^{-n-1}-2H_n(0)\right].
\end{equation}
This can be compared with the spectral coefficient $a_{N/2}(x|L_b)$,
related to the conformal anomaly by
$\left<T^\mu{}_\mu(x)\right>=a_{N/2}(x)(4\pi)^{N/2}$ \cite{BD82}. This
coefficient can be computed making an heat kernel expansion in the local
$\zeta$-function; it turns out that $a_{N/2}(x|A)=(4\pi)^{N/2}{\rm
Res}[\Gamma(s)\zeta(s,x|A)]_{s=0}$. This residue can be directly read in
equation (\ref{Zeven}), and coincides with the trace of the renormalized
stress tensor. We stress the fact that this last check, in the conformally
coupled case, is completely independent from the  $\zeta$-function
regularization.

\section{Application to various dimensions}\label{app}

In this Section, we shall apply the previous results to compute the
effective action, the field fluctuations and the stress tensor in various
dimensions.
In the even dimensional case we shall report only the results in $2$ and
$4$ dimensions, that can be compared with analogous expressions obtained
with other methods. Higher dimensional cases can be computed easily from
our general formulae; however, as the resulting expressions increase
considerably in complexity with the dimension, we shall omit them.\\ 
The odd dimensional case is simpler, and we shall give the expressions of
the effective action and the stress tensor up to $N=11$.

\subsection{$N=2$}\label{N2}
In two dimensions $\rho_2=1/2$, the coupling parameter is
$a^2b=1/4-2\xi+a^2m^2$ and $c_1^2=0$. The effective action (\ref{Ieff}) reads
\begin{eqnarray}
I_{eff}^{ {\mathbb H}^2}=-\frac1{2\pi a^2}\left[
\left(\frac18-\xi+\frac12a^2m^2\right)\left(1-\ln(a^2b)\right)
+2H'_0(0;a\sqrt b)
\right.\nonumber\\ 
\left.+\left(\frac16-\xi+\frac12a^2m^2\right)\ln(a^2\mu^2)\right],
\label{I2}
\end{eqnarray}
the expectation value of the field fluctuations (\ref{ffeven}) reads 
\begin{equation}
\left<\phi^2(x)\right>^{ {\mathbb H}^2}_{\mu^2}=-\frac1{2\pi}
\left[\psi\left( a\sqrt{b}+\frac12\right)-\ln(a\mu)\right],
\label{ff2}\end{equation} 
and the stress tensor (\ref{Teven}) is
\begin{equation}
\left<T_ {\mu\nu}(x)\right>^{ {\mathbb H}^2}_{\mu^2}=\frac1{4\pi a^2} \left[
\xi-\frac16-\frac12a^2m^2\left( 1-2\psi\left( a\sqrt{b}+\frac12\right)
+\ln(a^2\mu^2)\right)\right]\,g_ {\mu\nu}^{ {\mathbb H}^2}(x).
\label{stN2}\end{equation}

\subsection{$N=4$}
For a quantum scalar field propagating on $ {\mathbb H}^4$ we need the
coefficients $\rho_4=3/2$, $c^4_1=1/4$, $c^4_3=1$, $d_1=1$, $d_2=3/2$,
$H_0(0)=1/48$, $H_1(0)=7/1920$ and the functions (\ref{B01}) and
(\ref{B11}); the coupling parameter is $a^2b=9/4-12\xi+a^2m^2$. The
effective action is
\begin{eqnarray}
I_{eff}^{ {\mathbb H}^4}&=&\frac1{128\pi^2a^4}\left[
\frac{207}{16}+\frac{25}2a^2m^2+3a^4m^4-150\xi-72a^2m^2\xi+432\xi^2
\right.\nonumber\\ 
&&\left.-\left(\frac{63}8+8a^2m^2+2a^4m^4-96\xi-48a^2m^2\xi+288\xi^2\right)
\ln(b^2/\mu^2)-\frac{17}{60}\ln(a\mu)\right]\nonumber\\ 
&&+\frac1{8\pi^2a^4}\int_0^\infty\frac{\lambda\left(\lambda^2+\frac14
\right)}{e^{2\pi\lambda}+1}\ln\left(\lambda^2+a^2b\right)\,d\lambda ,
\end{eqnarray}
in agreement with \cite{Camporesi91}. The expectation value of the field
fluctuations (\ref{ffeven}) is given by
\begin{eqnarray}
\left<\phi^2(x)\right>_{\mu^2}^{ {\mathbb H}^4}&=&-\frac1{16\pi^2a^2}
\left(\frac73-12\xi+a^2m^2\right)\nonumber\\ 
&&-\frac1{8\pi^2a^2}\left[12\left(\xi-\frac16\right)-a^2m^2\right]\left[
\psi\left( a\sqrt b+\frac12\right)-\ln a\mu\right],
\end{eqnarray}
and the one loop renormalized stress tensor (\ref{Teven}) reads
\begin{eqnarray}
\left<T_ {\mu\nu}(x)\right>_{\mu^2}^{ {\mathbb H}^4}&=&
\frac1{32\pi^2a^4}\left\{
-\frac1{30}+36\left(\xi-\frac16\right)^2-12\left(\xi-\frac16\right)
a^2m^2+\frac16a^2m^2+\frac34a^4m^4\right.\nonumber\\ 
&&\left.-\left[a^4m^4-12a^2m^2\left(\xi-\frac16\right)\right]\left[
\psi\left( a\sqrt b+ {1\over 2}\right)-\ln(a\mu)\right]\right\}
g_{\mu\nu}^{{\mathbb H}^4}(x).
\end{eqnarray}
This tensor coincides with the known expression
\cite{Camporesi92}. Furthermore, for a conformally coupled field, $m=0$,
$a^2b=1/4$, the stress tensor is
\begin{equation}
\left<T_ {\mu\nu}(x)\right>_{conf}^{ {\mathbb H}^4}=-\frac1{960\pi^2a^4}
\,g_{\mu\nu}^{ {\mathbb H}^4}(x),
\end{equation}
and gives the correct conformal anomaly dictated by the spectral
coefficient $a_2(x|L_{conf})$.

\subsection{Odd Dimensions}
We report here the effective action and the stress tensor for odd
dimensions, up to $N=11$. We shall not
report the expressions of the field fluctuations, as in odd dimensions Eq.
(\ref{relffst}) reduces to a simple proportionality relation,
\begin{equation}
\left<T^\mu{}_\mu(x)\right>=-m^2\left<\phi^2(x)\right>.
\end{equation}
The effective actions are in accord with those computed in \cite{KB98}.

\subsubsection{$N=3$}
In three dimensions we have $\rho_3=1$, $a^2b=1-6\xi+a^2m^2$ and $c_2^3=1$;
the effective action reads
\begin{equation}
I_{eff}^{ {\mathbb H}^3}=\frac1{12\pi a^3}\left(1-6\xi+a^2m^2\right)^\frac32 ,
\end{equation}
and the expectation value of the stress tensor (\ref{Todd}) reads
\begin{equation}
\left<T_ {\mu\nu}(x)\right>_{\mu^2}^{ {\mathbb H}^3}=\frac{m^2}{12\pi a}
\sqrt{1-6\xi+a^2m^2}\,\,g_ {\mu\nu}^{ {\mathbb H}^3}(x).
\end{equation}

\subsubsection{$N=5$}
In five dimensions we have $\rho_5=2$, $a^2b=4-20\xi+m^2a^2$ and
$c_2^5=c_4^5=1$; the effective action reads
\begin{equation}
I_{eff}^{ {\mathbb H}^5}=-\frac1{360\pi^2a^5}\left(7-60\xi+3a^2m^2\right)
\left(4-20\xi+m^2a^2\right)^{3/2} ,
\end{equation}
and the expectation value of the stress tensor (\ref{Todd}) reads
\begin{equation}
\left<T_ {\mu\nu}(x)\right>
^{ {\mathbb H}^5}_{\mu^2}=-\frac{m^2}{120\pi^2a^3}\left(3-20\xi+a^2m^2
\right)\sqrt{4-20\xi+a^2m^2}\,\,g_ {\mu\nu}^{ {\mathbb H}^5}(x) .
\end{equation}

\subsubsection{$N=7$}
In seven dimensions $\rho_7=3$, $a^2b=9-42\xi+a^2m^2$, $c_2^7=4$, $c_4^7=5$
and $c_6^7=1$; the effective action reads
\begin{eqnarray}
I_{eff}^{ {\mathbb H}^7}&=&-\frac1{5040\pi^3a^7}\left(82+33m^2a^2+3m^4a^4
-1386\xi-252m^2a^2\xi+5292\xi^2\right)\nonumber\\ 
&&\times\left(9-42\xi+a^2m^2\right)^{3/2} ,
\end{eqnarray}
and the expectation value of the stress tensor (\ref{Todd}) reads
\begin{eqnarray}
\left<T_ {\mu\nu}(x)\right>^{ {\mathbb H}^7}_{\mu^2}&=&\frac{m^2}
{1680\pi^3a^5}\left(
40+13a^2m^2+a^4m^4-546\xi-84a^2m^2\xi+1764\xi^2
\right)\nonumber\\ 
&&\times\sqrt{9-42\xi+a^2m^2}\,\,g_ {\mu\nu}^{ {\mathbb H}^7}(x) .
\end{eqnarray}

\subsubsection{$N=9$}
In nine dimensions $\rho_9=4$, $a^2b=16-72\xi+a^2m^2$, $c_2^9=36$,
$c_4^9=49$, $c_6^9=14$ and $c_8^9=1$; the effective action reads
\begin{equation}
I_{eff}^{ {\mathbb H}^9}=-\frac1{151200\pi^4a^9}\left(-540+441a^2b
-90a^4b^2+5a^6b^3\right)
\left( a^2b\right)^{3/2} ,
\end{equation}
and the expectation value of the stress tensor (\ref{Todd}) reads
\begin{equation}
\left<T_ {\mu\nu}(x)\right>^{ {\mathbb H}^9}_{\mu^2}=
\frac{m^2}{30240\pi^4a^7}\left(
36-49a^2b+14a^4b^2-a^6b^3
\right)a \sqrt b\,\,g_ {\mu\nu}^{ {\mathbb H}^9}(x) .
\end{equation}

\subsubsection{$N=11$}
In eleven dimensions $\rho_{11}=5$, $a^2b=25-110\xi+a^2m^2$, $c_2^{11}=576$,
$c_4^{11}=820$, $c_6^{11}=273$, $c_8^{11}=30$ and $c_{10}^{11}=1$; the
effective action reads
\begin{equation}
I_{eff}^{ {\mathbb H}^{11}}=-\frac1{1995840\pi^5a^{11}}
\left(-6336+5412a^2b-1287a^4b^2+110a^6b^3-3a^8b^4\right)
\left( a^2b\right)^{3/2} ,
\end{equation}
and the expectation value of the stress tensor (\ref{Todd}) reads
\begin{equation}
\left<T_ {\mu\nu}(x)\right>^{ {\mathbb H}^{11}}_{\mu^2}=
\frac{m^2}{665280\pi^5a^9}\left(
576-820a^2b+273a^4b^2-30a^6b^3+a^8b^4
\right)a \sqrt b\,\,g_ {\mu\nu}^{ {\mathbb H}^{11}}(x) .
\end{equation}

\section{Conclusion}\label{concl}

In this paper we have obtained exact expressions at one loop for the
effective action and the vacuum expectation values of the field
fluctuations and the stress tensor for a scalar field propagating on an
$N$-dimensional hyperbolic space. Our expressions hold for massless as well
as massive fields, with an arbitrary coupling with the scalar curvature.\\ 
The computation of the stress tensor has been carried out with the recently
developed direct  $\zeta$-function  approach, which is known to be
equivalent to the point-splitting in compact spaces. Comparison of our
results with the known expressions in three and four dimensions sustains
the equivalence of the $\zeta$-function  and point-splitting approaches
also in presence of a continuous spectrum.\\
The computation presented here is the first step to the study of physically
more interesting cases. Making use of Selberg-like trace formulae to extend
this work to quotient spaces $ {\mathbb H}^N/\Gamma$, it is possible to
investigate finite temperature effects on AdS spacetime, and quantum
corrections to the metric and entropy of the higher dimensional constant
curvature black holes.

\section*{acknowledgments}
I would like to thank Valter~Moretti, Luciano~Vanzo and Sergio~Zerbini for
useful discussions and for reading the manuscript.

\appendix
\section{The Function $H_n(s;\mu)$}

In this section we shall study the integral $H_n(s;\mu)$ defined in
Eq. (\ref{Bn}). It defines an analytic function on the whole complex
$s$-plane. We are interested in the values it takes, with its derivative,
in $s=0$ and $s=1$.\\ 
The integral can be exactly computed in $s=0$ using Eq. (3.411.4) of \cite{GR}
\begin{equation}
H_n(0;\mu)=\left(1-2^{-2n-1}\right)\frac{|B_{2n+2}|}{4n+4},
\end{equation}
where the $B_n$ are the Bernoulli numbers. In $s=1$, we have
\begin{equation}
  H_0(1;\mu)= {1\over 2}\,\psi\left( \mu+ {1\over 2}\right)-\frac14\ln(\mu^2);
\label{B01}\end{equation}
to evaluate $H_n(1;\mu)$ in $n=1,2,\dots$, we start from the identity
\begin{equation}
\int_0^\infty\frac{x^{2n+1}}{e^{2\pi x}+1}\ln(\alpha
x^2+\mu^2)\,dx=\ln\alpha\int_0^\infty\frac{x^{2n+1}\,dx}{e^{2\pi x}+1}
+\int_0^\infty\frac{x^{2n+1}}{e^{2\pi x}+1}\ln(x^2+\mu^2/\alpha)\,dx.
\end{equation}
Taking the derivative with respect to $\alpha$ and setting $\alpha=1$, one
obtains the recurrence relation
\begin{equation}
H_{n+1}(1;\mu)=X_{n+1}-\mu^2 H_n(1;\mu),
\label{recursion}\end{equation}
where we have defined
\begin{equation}
X_n=(1-2^{1-2n})\frac{|B_{2n}|}{4n}.
\end{equation}
From equation (\ref{recursion}) one finds finally
\begin{equation}
H_n(1;\mu)=\sum_{k=1}^nX_k(-\mu^2)^{n-k}+(-\mu^2)^nH_0(1;\mu).
\label{Bn1}\end{equation}
In particular we shall need the case $n=1$, for which,
\begin{equation}
  H_1(1;\mu)=\frac1{48}- {1\over 2} \mu^2\,\,\psi\left(\mu+ 
    {1\over2}\right)+\frac14\mu^2\ln(\mu^2). 
\label{B11}\end{equation}
Let us turn now to the derivative of $H_n(s;\mu)$; let 
\begin{equation}
H_n'(0;\mu)=-\int_0^\infty\frac{\lambda^{2n+1}\ln\left(\lambda^2+\mu^2\right)}
{e^{2\pi\lambda}+1}\,d\lambda;
\end{equation}
integrating in the variable $\mu^2$ the relation 
\begin{equation}
\frac{\partial H'_n(0;\mu)}{\partial(\mu^2)} = -H_n(1;\mu),
\end{equation}
using (\ref{Bn1}), one obtains
\begin{eqnarray}
H'_n(0;\mu)&=&H'_n(0;0)+\sum_{k=1}^{n}X_k\frac{(-\mu^2)^{n-k+1}}{n-k+1}+
(-1)^n\mu^{2n+2}\left[\frac{\ln\mu}{2n+2}-\frac1{4(n+1)^2}\right]\nonumber\\ 
&&+(-1)^{n+1}\int_0^\mu\mu^{2n+1}\psi\left(\mu+\frac12\right)\,d\mu,
\label{a}\end{eqnarray}
where $H'_n(0;0)$ is the constant
\begin{eqnarray}
H'_n(0;0)&=&-2\int_0^\infty\frac{\lambda^{2n+1}\ln\lambda}
{e^{2\pi\lambda}+1}\,d\lambda\nonumber\\ 
&=&-2\frac{(2n+1)!}{(2\pi)^{2n+2}}(1-2^{-2n-1})\zeta'_R(2n+2)\nonumber\\
&&-\left[(1-2^{-2n-1})(d_{2n+1}-\gamma-\ln 2\pi)+2^{-2n-1}\ln 2\right]
\frac{|B_{2n+2}|}{2n+2}.
\end{eqnarray}
Note that the integral in Eq. (\ref{a}) can evaluated in terms of multi-gamma
functions (see Appendix of \cite{KB98}).

\end{document}